\documentclass[prb,aps,twocolumn,showpacs,superscriptaddress]{revtex4}
\usepackage{graphicx,amsmath,amsfonts,latexsym,bm}

  \def\eq{&=&} \def\nn{\nonumber}

\begin{document} 
\title{Magnetic screening properties of superconductor-ferromagnet
  bilayers}

\author{Manuel Houzet}

\affiliation{CEA, INAC, SPSMS, F-38054 Grenoble, France}

\author{Julia S. Meyer}

\affiliation{Department of Physics, The Ohio State University,
  Columbus, Ohio 43210, USA}

\affiliation{Universit\'e Joseph Fourier, F-38041 Grenoble, France}
 
\date{\today}
 
% 74.25.-q Properties of type I and type II superconductors 74.25.Ha
% Magnetic properties 74.25.Nf Response to electromagnetic fields
% (nuclear magnetic resonance, surface impedance, etc.)  74.45.+c
% Proximity effects; Andreev effect; SN and SNS junctions 74.78.-w
% Superconducting films and low-dimensional structures 74.78.Fk
% Multilayers, superlattices, heterostructures

\begin{abstract}
  We study theoretically the magnetic screening properties of thin,
  diffusive superconductor/ferromagnet bilayers subject to a
  perpendicular magnetic field.  We find that the effective
  penetration depth characterizing the magnetic response oscillates
  with the thickness of the ferromagnetic layer on the scale of the
  ferromagnetic coherence length.
\end{abstract}
 
\maketitle
 
While superconductor-normal metal (SN) structures have been
intensively studied for decades, superconductor-ferromagnet (SF)
structures have only become accessible recently because of the much
reduced length scales in ferromagnets. Due to their incompatible spin
properties, the proximity effect between a singlet superconductor and
a ferromagnet leads to a variety of unusual
phenomena~\cite{Buzdin2005,Bergeret2005}.  Through the exchange field
acting on the electron spins in the ferromagnet, Cooper pairs acquire
a finite momentum $\delta k$ which leads to an oscillatory behavior of
the anomalous Green function~\cite{Demler1997}. Observable
consequences are, e.g., a non-monotonic dependence of the transition
temperature~\cite{Buzdin1990,Radovic1991,Jiang1995} and the density of
states at the Fermi level~\cite{Buzdin2000,Kontos2001} on the
thickness of the F layer in SF bilayers, and the possibility of
$\pi$-Josephson junctions at certain thicknesses of the F layer in
superconductor-ferromagnet-superconductor (SFS)
trilayers~\cite{Buzdin1982,Buzdin1991,Ryazanov2001PRL}.

While most experiments on hybrid systems use resistive measurements,
screening of an external magnetic field offers an alternative tool to
study the proximity effect. These measurements probe deeply into the
superconducting state because they provide both the magnitude and
temperature dependence of the effective superfluid density.  Various
configurations for the magnetic response can be considered.  The
magnetization of SN hybrids with a magnetic field applied parallel to
their interface has been addressed theoretically in
Ref.~\cite{Zaikin1982}, with still debated experimental results in the
case of SN cylinders~\cite{Visani1990,Bernd2000}.  Alternatively, the
screening properties of thin films can be probed by measuring the
mutual inductance of two coils positioned on opposite sides of the
sample~\cite{Turneaure1996,Turneaure1998}. The mutual inductance can
be related to the complex conductivity of the film which in turn can
be related to the screening length $\lambda$ or the superfluid density
$\rho_S$. To be precise, in SF bilayers, one measures the superfluid
density $\rho_S \propto \lambda^{-2}$ integrated over the width of the
bilayer or an effective screening length
\begin{eqnarray}
  \lambda_{\rm eff}^{-2} = d_S^{-1}\int_{-d_S}^{d_F} dx \;\lambda^{-2}(x),
\end{eqnarray}
where $d_S$ and $d_F$ are the thicknesses of the superconducting and
ferromagnetic layer, respectively, and $x$ is the coordinate normal to
the interface. First experimental results on Nb/Ni bilayers have been
reported using this setup in Ref.~\cite{Lemberger2008}, where a
non-monotonic dependence of the effective screening length
$\lambda_{\rm eff}^{-2}$ on the thickness of the Ni layer has been
observed.

In this paper, we study the screening length $\lambda_{\rm eff}$ of a
SF bilayer subject to a weak perpendicular magnetic field. The main
assumptions are that {\em (i)} the exchange field $h$ in the
ferromagnet is much larger than the superconducting order parameter
$\Delta$, {\em (ii)} the system is in the dirty limit and, thus, the
Usadel equation~\cite{Usadel1970} can be used, {\em (iii)} the
screening length $\lambda_{\rm eff}$ is much larger than the thickness
$d=d_S+d_F$ of the bilayer, and {\em (iv)} the width $d_S$ of the
superconducting (S) layer is smaller than the superconducting
coherence length $\xi_S=\sqrt{D_S/(2\pi T_{c0})}$, where $D_S$ is the
diffusion constant and $T_{c0}$ is the transition temperature of the
bare S layer. Our main result is that the screening length displays an
oscillatory behavior with the thickness of the ferromagnet.

Due to the normalization condition $\hat g^2=1$ of the quasi-classical
Usadel Green function $\hat g$, it can be parametrized by an angle
$\theta$ such that the normal Green function $G=\cos\theta$ whereas
the anomalous Green function $F=\sin\theta$.  The system is then
described by four coupled equations in terms of the angles $\theta_S$
on the S side of the SF interface, $\theta_0$ on the ferromagnetic (F)
side of the SF interface, and $\theta_F$ at the ferromagnet-vacuum
interface.

The Usadel equation of the F layer,
$-D_F\nabla^2\theta+2ih\sin\theta=0$, can be integrated to yield
\begin{eqnarray}
  2\sqrt{i}y\eq\int_{\theta_F}^{\theta_0}d\theta\frac1{\sqrt{\cos\theta_F-\cos\theta}},\label{eq-F}
\end{eqnarray}
where $y=d_F/\xi_F$ and $\xi_F=\sqrt{D_F/h}$ is the ferromagnetic
coherence length, with the diffusion constant $D_F$ of the F
layer. The boundary condition imposing current conservation at the SF
interface~\cite{Kupriyanov1988} can be expressed as
\begin{eqnarray}
  \sin(\theta_S-\theta_0)\eq 2\sqrt{i}\beta\sqrt{\cos\theta_F-\cos\theta_0},\label{eq-bc}
\end{eqnarray}
where $\beta=R_b\sigma_F/{\xi_F}$. Here $R_b$ is the interface
resistance per square, and $\sigma_F$ is the conductivity of the F
layer. In the limit $d_S\ll\xi_S$, the Usadel equation of the S layer,
$-D_S\nabla^2\theta+2\omega\sin\theta=2\Delta\cos\theta$, where
$\omega$ is a fermionic Matsubara frequency, can be simplified by an
expansion in small spatial variations of the angle $\theta$ across the
S layer combined with the boundary condition (\ref{eq-bc}). One
obtains
\begin{eqnarray}
  \omega\sin\theta_S+2\sqrt{i}\alpha\sqrt{\cos\theta_F-\cos\theta_0}\eq\Delta\cos\theta_S,\label{eq-S}
\end{eqnarray}
where $\alpha={D_S\sigma_F}/({2\sigma_Sd_S\xi_F})$ and $\sigma_S$ is
the conductivity of the S layer.  Finally, the self-consistency
equation for the order parameter $\Delta$ reads
\begin{eqnarray}
  \Delta\eq\pi T\lambda_{\rm BCS}\Re\left[\sum_\omega\sin\theta_S\right],\label{eq-gap}
\end{eqnarray}
where $\lambda_{\rm BCS}$ is the BCS coupling constant.

In diffusive superconductors, the screening length $\lambda$ describes
the local (London) current response~\cite{degennes} to a
vector-potential, $\bm{j}=-1/(\mu_0\lambda^2)\bm{A}$, where
$\lambda^{-2}=(2\pi T\mu_0\sigma_S/\hbar )\sum_\omega F^2$ is
proportional to the superfluid density. Here $\mu_0$ is the vacuum
permeability. In SF bilayers, the effective screening length is
related to the angles $\theta$ through the equation
\begin{eqnarray}
  \frac1{\lambda_{\rm eff}^2}=\frac{2\pi T\mu_0\sigma_S}{\hbar}\Re\left[\sum_\omega\big(\sin^2\theta_S+\gamma\!\int\limits_0^{y}\!dx\,\sin^2\theta(x)\big)\right]\!\!,
\end{eqnarray}
% \end{widetext}
where $\gamma=\sigma_F\xi_F/(\sigma_Sd_S)$. Using the Usadel equation
of the F layer, the integral over $x$ can be traded for an integral
over $\theta$, namely $dx=\frac1{2\sqrt
  i}(\cos\theta_F-\cos\theta)^{-1/2}d\theta$, ranging from $\theta_F$
to $\theta_0$.

In general the set of equations (\ref{eq-F})-(\ref{eq-gap}) can be
solved numerically only, but in some limiting cases an analytical
solution is possible.  At $T=0$, a simplification occurs because the
sums over $\omega$ can be replaced by integrals, and subsequently the
integration over $\omega$ can be traded for an integration over
$\theta_S$ using the Usadel equation~\cite{baladie-buzdin}. It is then
sufficient to solve the Usadel equation at $\omega=0$ for
$\theta_S(0)$. In particular, using Eqs.~(\ref{eq-F}) and
(\ref{eq-bc}), the Usadel equation of the S layer (\ref{eq-S}) can be
brought into the form
$(\omega+F(\theta_S))\sin\theta_S=\Delta\cos\theta_S$, yielding
$d\omega=-(\Delta/\sin^2\theta_S+F'(\theta_S))d\theta_S$. Using this
trick, the zero temperature gap $\Delta$ is given as
\begin{eqnarray}
  \ln\frac\Delta{\Delta_0}=\Re\left[\ln\tan\frac{\theta_S(0)}2+\int\limits_0^{\theta_S(0)}\!\!d\theta_S\,F'(\theta_S)\sin\theta_S\right],
\end{eqnarray}
where $\Delta_0$ is the zero-temperature gap of the bare S layer, a
result which can then be used to compute $\lambda_{\rm eff}^{-2}$.

In the following, we provide analytical results for the effective
screening length in two limits, namely {\em (i)} for a system without
barrier $\beta=0$ and {\em (ii)} for a system with a strong barrier
$\beta\gg1$. In both cases, solutions are presented for small
parameters $\alpha$. For convenience, we introduce the notation
$\tilde x=(1+i)x$ for $x=\alpha,\beta,y$.

In the absence of a barrier $\beta=0$, the boundary condition
(\ref{eq-bc}) imposes that the angles on both sides of the SF
interface, $\theta_0$ and $\theta_S$, are the same.

If $d_F\gg\xi_F$, the angle $\theta_F$ is small, and Eq.~(\ref{eq-F})
yields $\theta_F=8\tan\frac{\theta_S}4\exp[-\tilde y]$. Thus, we can
simplify Eq.~(\ref{eq-S}) to yield
\begin{eqnarray}
  \omega\sin\theta_S+2\tilde\alpha\sin\frac{\theta_S}2\left(1-8e^{-2\tilde y}\frac{\tan^2\frac{\theta_S}4}{\sin^2\frac{\theta_S}2}\right)=\Delta\cos\theta_S.
\end{eqnarray}
Treating $\alpha\ll\Delta_0$ perturbatively, one finds
$\theta_S(0)=\frac\pi2+\delta\theta_S$, where
$\delta\theta_S=-\sqrt2(\tilde\alpha/\Delta_0)\left(1-16e^{-2\tilde
    y}(3-2\sqrt2)\right)$, and
\begin{eqnarray}
  \delta\Delta\eq-2\alpha\left(\sqrt2-\ln(1+\sqrt2)\right)\\
  &&-\frac{16}3\Re\left[\tilde\alpha e^{-2\tilde y}\right]
  % \alpha e^{-2y}(\cos(2y)+\sin(2y))
  \left(3\ln(1+\sqrt{2})+4-5\sqrt2\right).\nn
\end{eqnarray}
This solution describes gapless superconductivity with a finite
density of states $\nu(0)$ at the Fermi level in the superconductor
that oscillates with the thickness of the ferromagnet:
$\nu(0)=-\nu_0\Re[\delta \theta_S]$, where $\nu_0$ is the density of
states in normal state.  The equation for the screening length takes
the form
\begin{eqnarray}
  \!\!\!\!\!\frac{\lambda_0^2}{\lambda_{\rm eff}^2(0)}\eq1-\left(a_1-a_2(\cos 2y+\sin 2y)e^{-2y}\right)\frac\alpha{\Delta_0}\label{eq-lam_thick1}
  \\
  &&
  +\frac{2\sqrt2}{3\pi}\gamma\left(1+a_3 \,y\,e^{-2y}\cos 2y-a_4\frac\alpha{\Delta_0}\right),
  \nonumber
\end{eqnarray}
where $\lambda_0^{-2}=\pi\mu_0\sigma_S\Delta_0/\hbar$ is the inverse
screening length of the bare S layer at zero temperature, and the
coefficients $a_i$ are positive~\cite{coefficients}. Both, the
contributions to the effective screening length from the S layer and
from the F layer ($\propto\gamma$), oscillate on the length scale of
the ferromagnetic coherence length.
 
If on the other hand $d_F\ll\xi_F$, the variation of the angle
$\theta$ is small across the F layer and, thus,
$\theta_S-\theta_F\ll1$. Using Eq.~(\ref{eq-F}), one obtains
$\theta_S=\theta_F\cosh\tilde y$.  Inserting this relation into the
Usadel equation of the S layer results in
\begin{eqnarray}
  \left(\omega+2i\alpha y+\frac43 \alpha y^3\cos\theta_S\right)\sin\theta_S\eq\Delta\cos\theta_S.\label{eq-S_y0}
\end{eqnarray}
We find $\delta\theta_S=-2i\alpha y/\Delta_0$ and
$\delta\Delta=-\frac\pi3\alpha y^3$.  Note that because
$\theta_S(0)=\frac\pi2+i\phi$, where $\phi$ real, the density of
states possesses a gap in this regime.  Using Eq.~(\ref{eq-S_y0}) to
convert the integral over $\omega$ to an integral over $\theta_S$, the
screening length is given by
\begin{eqnarray}
  \frac{\lambda_0^2}{\lambda_{\rm eff}^2(0)}\eq1-\frac\pi3(1+\frac{16}{3\pi^2})\frac\alpha{\Delta_0}y^3+\gamma y.\label{eq-lam_thin1}
\end{eqnarray}
Eq.~(\ref{eq-lam_thin1}) predicts an increase in $\lambda_{\rm
  eff}^{-2}$ as long as $d_F<\xi_F^2/\xi_S$ before it starts to
decrease. The regime $d_F\sim\xi_F$ connecting the results
Eq.~(\ref{eq-lam_thick1}) and (\ref{eq-lam_thin1}) is treated
numerically (see below).

In the opposite limit of a strong barrier, $\beta\gg1$, both
$\theta_0$ and $\theta_F$ are small, if the F layer is not too thin,
$y\gg\beta^{-1}$. Eq.~(\ref{eq-F}) then yields
$\theta_0=\theta_F\cosh\tilde y$, and, using the boundary condition,
the Usadel equation of the S layer can be rewritten as
\begin{eqnarray}
  \left(\omega+\frac\alpha\beta-\frac\alpha{\sqrt{2i}\beta^2\tanh\tilde y}\cos\theta_S\right)\sin\theta_S=\Delta\cos\theta_S.\label{eq-bl_th}
\end{eqnarray}
Eq.~(\ref{eq-bl_th}) yields $\delta\theta_S=-\alpha/(\beta\Delta_0)$,
and thus no gap in the density of states,
$\nu(0)=\nu_0\alpha/(\beta\Delta_0)$, while
\begin{eqnarray}
  \delta\Delta=-\frac\alpha\beta+\frac{\pi\alpha}{8\beta^2}\Re\left[(1-i)\coth\tilde y\right].
\end{eqnarray}
The screening length is given as
\begin{widetext}
  \begin{eqnarray}
    \frac{\lambda_0^2}{\lambda_{\rm eff}^2(0)}\eq 1-(1+\frac2\pi)\frac\alpha{\beta\Delta_0}+(\frac\pi8+\frac2{3\pi})\frac{\alpha}{\beta^2\Delta_0}\Re\left[(1-i)\coth\tilde y\right]-\frac{\gamma}{16\beta^2}\Re\left[\frac{4iy+(1+i)\sinh2\tilde y}{\sinh^2\tilde y}\right].
  \end{eqnarray}
\end{widetext}
Again the effective screening length oscillates on the scale of
$\xi_F$. However, these oscillations are suppressed due to the large
barrier. For $y\gg1$, the oscillatory function in the contribution of
the S layer has the same form as the one in (\ref{eq-lam_thick1})
whereas the oscillating part of the contribution from the F layer is
proportional to $y\,e^{-2y}\,\sin(2y)$.

In the case of a very thin F layer, $y\ll\beta^{-1}$, the variation of
the angle $\theta$ is small across the F layer, see above.  The
boundary condition at the SF interface then simplifies, and the Usadel
equation of the S layer yields
\begin{eqnarray}
  \left(\omega+2i\alpha y+4\alpha\beta y^2\cos\theta_S\right)\sin\theta_S\eq\Delta\cos\theta_S.\label{eq-S_y0-2}
\end{eqnarray}
We find $\delta\theta_S=-2i\alpha y/\Delta_0$ and
$\delta\Delta=-\pi\alpha\beta y^2$. As for the case without a barrier,
the density of states in the thin film regime is gapped.  Using
Eq.~(\ref{eq-S_y0-2}) to convert the integral over $\omega$ to an
integral over $\theta_S$, the screening length is given by
\begin{eqnarray}
  \frac{\lambda_0^2}{\lambda_{\rm eff}^2(0)}\eq 1-\pi(1+\frac{16}{3\pi^2})\frac{\alpha\beta}{\Delta_0}y^2+\gamma y.
\end{eqnarray}
The inverse screening length increases in the very narrow regime
$d_F<\xi_F^3/(\beta \xi_S^2)$.

Thus, we find oscillations of the screening length both in the absence
of a barrier and in the presence of a strong barrier. The amplitude of
oscillations in the latter case is suppressed, however. In both cases,
analytic results can be found for small thicknesses $y<y^*$ and large
thicknesses $y>y^*$, where $y^*\sim\min\{1,\beta^{-1}\}$ denotes the
position of the first strong minimum. The vicinity of this minimum is
not accessible to analytic solution.

\begin{figure}[h!]
  \centerline{\includegraphics[width=0.4\textwidth]{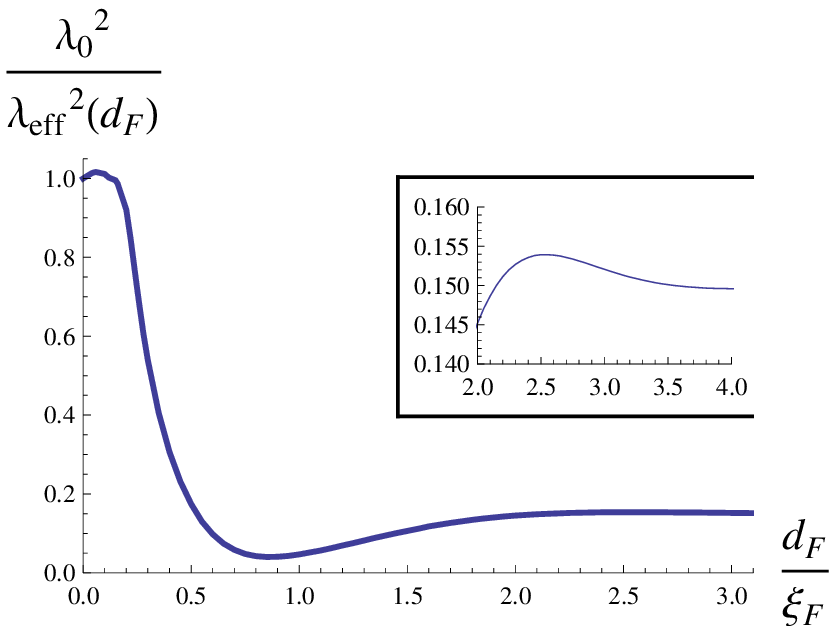}}
  \caption{ Oscillation of the inverse screening length
    $1/\lambda_{\rm eff}^2$ as a function of $d_F$ at temperature
    $T=0.1T_{c0}$. Here we use the parameters $\alpha=1.2$, $\beta=1$,
    and $\gamma=0.6$. The inset magnifies the weak maximum at
    $d_F\approx2.5\xi_F$.}
  \label{fig1}
\end{figure}

To find a solution in this regime, we note that Eq.~(\ref{eq-F})
yields a general relation between $\theta_0$ and $\theta_F$, namely
\begin{eqnarray}
  \sin\frac{\theta_0}2\eq\frac1{\sqrt{1+\cot^2\frac{\theta_F}2{\rm
        \,cn\,}^2\left(\tilde y,\cos^2\frac{\theta_F}2\right)}},\label{eq-cn}
\end{eqnarray}
where cn is the Jacobi elliptic function. Using Eq.~(\ref{eq-cn}), the
boundary condition (\ref{eq-bc}) yields $\theta_S$ as a function of
$\theta_F$. Inserting this solution into the other equations, the set
of coupled Eqs.~(\ref{eq-S}) and (\ref{eq-gap}) can then be solved
numerically.  The thickness dependence of the low-temperature
screening length is displayed in Fig.~\ref{fig1}. The minimum at
$d_F\sim\xi_F$ is clearly visible whereas further oscillations at
larger $d_F$ are very small.

\begin{figure}[h!]
  \centerline{\includegraphics[width=0.45\textwidth]{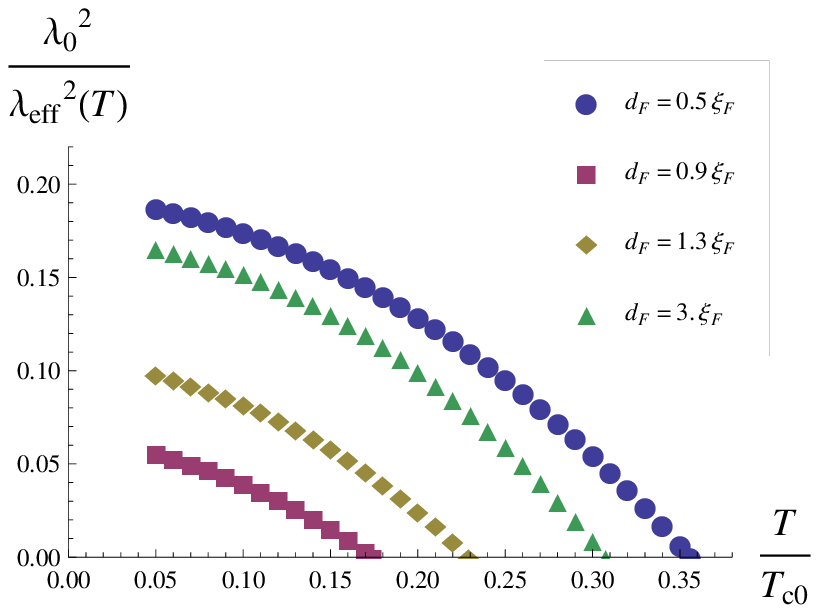}}
  \caption{Temperature dependence of $\lambda_{\rm eff}^{-2}$. The
    parameters used are the same as in Fig.~\ref{fig2}, and four
    different thicknesses are shown: $d_F/\xi_F=$ 0.5, 0.9, 1.3, and
    3. The corresponding critical temperatures are
    $T_c(0.5)=0.36T_{c0}$, $T_c(0.9)=0.17T_{c0}$ (close to the
    minimum), $T_c(1.3)=0.23T_{c0}$, and $T_c(0.5)=0.31T_{c0}$ (close
    to the asymptotic value for $d_F/\xi_F\gg1$).  }
  \label{fig2}
\end{figure}

Furthermore, the numeric solution allows one to describe the
temperature dependence of the screening length. Fig.~\ref{fig2} shows
the finite temperature curves for different values of $d_F/\xi_F$. The
oscillations of the zero-temperature screening length mirror the
oscillations of the critical temperature as well as the slope of the
screening length close to $T_c$.

In the vicinity of the critical temperature $T_c$ an analytic solution
is possible for all parameter values. For the SF bilayer, the critical
temperature is given by the solution of the equation~\cite{Buzdin2005}
\begin{eqnarray}
  \ln\frac
  {T_c}{T_{c0}}=\Psi\left(\frac12\right)-\Re\left[\Psi\left(\frac12+\frac1{2\pi T_c\tau_s}\right)\right],\label{eq-T_c}
\end{eqnarray}
where the (complex) relaxation time $\tau_s$ reads
\begin{eqnarray}
  \tau_s^{-1}\eq\frac{\sqrt{2i}\,\alpha\tanh(\sqrt{2i}\,y)}{1+\sqrt{2i}\,\beta\tanh(\sqrt{2i}\,y)}.\label{eq-tau_s}
\end{eqnarray}
Eqs.~(\ref{eq-T_c},\ref{eq-tau_s}) are obtained by linearizing
(\ref{eq-F})-(\ref{eq-gap}) in small $\theta$ close to the
transition. For $\tau_s^{-1}$ small,
$T_c=T_{c0}-\frac\pi4\Re\left[\tau_s^{-1}\right]$. For
$|\tau_s^{-1}|=\Delta_0/2$, the transition temperature vanishes
according to Eq.~(\ref{eq-T_c}). Note, however, that for large
$\tau_s^{-1}$ the transition typically becomes first
order~\cite{Tollis2004}.

Expansion of equations (\ref{eq-F})-(\ref{eq-gap}) up to cubic order
then yields the temperature dependence of the screening length
$\lambda_{\rm eff}$ close to $T_c$. Namely,
\begin{widetext}
  \begin{eqnarray}
    \frac1{\lambda_{\rm eff}^2(T)}\eq\frac{\mu_0\sigma_S}{\hbar\pi
      T_c}\Delta^2(T)\Re\left[\left(1+\frac\gamma{4\sqrt{2i}}\frac{2\tilde
          y+\sinh(2\tilde y)}{(\cosh\tilde y+\tilde\beta\sinh\tilde y)^2}\right)\Psi^{(1)}\left(\frac12+\frac1{2\pi
          T_c\tau_s}\right)\right].
  \end{eqnarray}
\end{widetext}
We see that the contribution of the F layer to $\lambda_{\rm
  eff}^{-2}$ displays an oscillatory dependence on its
thickness. Furthermore, both $T_c$ and $\Delta(T)$ oscillate. In
particular,
% \begin{widetext}
\begin{eqnarray}
  \frac{\Delta^2(T)}{T_c-T}
  =
  \frac{4\pi T_c\left(1-\Re\left[(2\pi T_c\tau_s)^{-1}\Psi^{(1)}(z)\right]\right)}{-\Re\left[\Psi^{(2)}(z)-f(\tau_s,\beta,y)\Psi^{(3)}(z)\right]},
  % \nonumber \\
\end{eqnarray}
% \end{widetext}
at $T<T_c$, where $z=\frac12+(2\pi T_c\tau_s)^{-1}$ and
\begin{eqnarray}
  f(\tau_s,\beta,y)\eq\frac1{48}(2\pi T_c\tau_s)^{-1}\frac1{(1+\tilde\beta\tanh\tilde
    y)^3}
  \\
  & &
  \times
  \left(4(1+2\tilde\beta\tanh\tilde y)^2-\frac{2\tilde y+\sinh(2\tilde y)}{\sinh\tilde y\cosh^3\tilde y}\right).
  \nonumber
\end{eqnarray}
Note that the simple relation $\lambda^{-2}\propto\Delta^2$ does not
hold in the presence of the F layer. The slope of $\lambda_{\rm
  eff}^{-2}$ close to $T_c$ has its own dependence on the thickness of
the F layer and the relaxation rate $\tau_s^{-1}$.

In conclusion, we have shown that the screening length of SF bilayers
displays a oscillatory dependence on the thickness of the F
layer. Analytic solutions have been found in various regimes and a
general solution has been determined numerically. The obtained
non-monotonic dependence of the screening length has been observed
experimentally~\cite{Lemberger2008}. Our method can be easily extended
to more complicated situations such as multilayers where unusual
features of the proximity effect have been
predicted~\cite{Buzdin2005,Bergeret2005}.

\begin{acknowledgments} 
  We would like to acknowledge A. Buzdin and T. Lemberger for many
  discussions.  JSM thanks the CEA/INAC/SPSMS for hospitality.
\end{acknowledgments}


\begin{thebibliography}{}

\bibitem{Buzdin2005} A. I. Buzdin,
  % {\em Proximity Effects in Superconductor-Ferromagnet
  %   Heterostructures},
  Rev. Mod. Phys. {\bf 77}, 935 (2005).
  
\bibitem{Bergeret2005} F. S. Bergeret, A. F. Volkov, and K. B. Efetov,
  % {\em Odd Triplet Superconductivity and Related Phenomena in
  %   Superconductor-Ferromagnet Structures},
  Rev. Mod. Phys. {\bf 77}, 1321 (2005).
 
  
\bibitem{Demler1997} E. A. Demler, G. B. Arnold, and M. R. Beasley,
  % {\em Superconducting Proximity Effects in Magnetic Metals},
  Phys.  Rev. B {\bf 55}, 15174 (1997).
  
\bibitem{Buzdin1990} A. I. Buzdin and M. Yu. Kupriyanov,
  % {\em Transition-Temperature of a Superconductor-Ferromagnet
  %   Superlattice},
  JETP Lett. {\bf 52}, 487 (1990).
  
\bibitem{Radovic1991} Z. Radovic, M. Ledvij, L. Dobrosavljevic-Grujic,
  A. I. Buzdin, and J. R. Clem,
  % {\em Transition-Temperatures of Superconductor-Ferromagnet
  %   Superlattices},
  Phys. Rev. B {\bf 44}, 759 (1991).
  
\bibitem{Jiang1995} J. S. Jiang, D. Davidovic, D. H. Reich, and
  C. L. Chien,
  % {\em Oscillatory Superconducting Transition Temperature in Nb/Gd
  %   Multilayers},
  Phys. Rev. Lett. {\bf 74}, 314 (1995).

\bibitem{Buzdin2000} A. I. Buzdin,
  % {\em Density of states oscillations in a ferromagnetic metal
  %   in contact with a superconductor},
  Phys. Rev. B {\bf 62}, 11377 (2000).
	
\bibitem{Kontos2001} T. Kontos, M. Aprili, J. Lesueur J, and
  X. Grison,
  % {\em Inhomogeneous Superconductivity Induced in a Ferromagnet by
  %   Proximity Effect},
  Phys. Rev. Lett. {\bf 86}, 304 (2001).
  

\bibitem{Buzdin1982} A. I. Buzdin, L. N. Bulaevskii, and S. V.
  Panyukov,
  % {\em Critical-Current Oscillations as a Function of the Exchange
  %   Field and Thickness of the Ferromagnetic Metal (F) in and S-F-S
  %   Josephson Junction},
  JETP Lett. {\bf 35}, 178 (1982).
  
\bibitem{Buzdin1991} A. I. Buzdin and M. Yu. Kupriyanov,
  % {\em Josephson Junction with a ferromagnetic layer},
  JETP Lett. {\bf 53}, 321 (1991).
  
\bibitem{Ryazanov2001PRL} V. V. Ryazanov, V. A. Oboznov,
  A. Yu. Rusanov, A. V.  Veretennikov, A. A. Golubov, and J. Aarts,
  % {\em Coupling of Two Superconductors through a Ferromagnet:
  %   Evidence for a $\pi$-Junction},
  Phys. Rev. Lett. {\bf 86}, 2427 (2001).

\bibitem{Zaikin1982} A. D. Zaikin,
  % {\em Meissner Effect in Superconductor Normal Metal Proximity
  %   Sandwiches},
  Solid State Comm. {\bf 41}, 533 (1982).

\bibitem{Visani1990} P. Visani, A. C. Mota, and A. Pollini,
  % {\em Novel reentrant effect in the proximity-induced
  %   superconducting behavior of silver},
  Phys.  Rev. Lett {\bf 65}, 1514 (1990).

\bibitem{Bernd2000} F. Bernd M\"uller-Alinger and A. C. Mota,
  % {\em Paramagnetic reentrant effect in high purity mesoscopic AgNB
  %   proximity structures},
  Phys.  Rev. Lett {\bf 84}, 3161 (2000).

\bibitem{Turneaure1996} S. J. Turneaure, E. R. Ulm, and T. R.
  Lemberger,
  % {\em Numerical Modeling of a Two-Coil Apparatus for Measuring the
  %   Magnetic Penetration Depth in Superconducting Films and Arrays},
  J.  Appl. Phys. {\bf 79}, 4221 (1996).
  
\bibitem{Turneaure1998} S. J. Turneaure, A. Pesetski, and T. R.
  Lemberger,
  % {\em Numerical Modeling and Experimental Considerations for a
  %   Two-Coil Apparatus to Measure the Complex Conductivity of
  %   Superconducting Films},
  J. Appl. Phys. {\bf 83}, 4334 (1998).

\bibitem{Lemberger2008} T. R. Lemberger, I. Hetel, A. J. Hauser, and
  F. Y.  Yang,
  % {\em Superfluid Density of Superconductor-Ferromagnet Bilayers},
  J. Appl. Phys. {\bf 103}, 07C701 (2008).


\bibitem{Usadel1970} K. D. Usadel,
  % {\em Generalized Diffusion Equation for Superconducting Alloys},
  Phys. Rev. Lett. {\bf 25}, 507 (1970).

\bibitem{Kupriyanov1988} M. Y. Kupriyanov and V. F. Lukichev,
  % {\em Effect of Boundary Transparency on Critical Current in Dirty
  %   SS'S Structures},
  Sov.  Phys.  JETP {\bf 67}, 1163 (1988).
  
\bibitem{degennes} See e.g.  P. G. De Gennes, {\em Superconductivity
    Of Metals And Alloys} (Addison Wesley, 1989).

\bibitem{baladie-buzdin} I. Baladi\'{e} and A. I. Buzdin,
  % {\em Thermodynamic properties of
  %   ferromagnet/superconductor/ferromagnet nanostructures}
  Phys. Rev. B {\bf 67}, 014523 (2003).

\bibitem{coefficients} The numerical values of the coefficients are
  $a_1=1.77$, $a_2=3.82$, $a_3=6.44$, and $a_4=3.19$.

\bibitem{Tollis2004} S. Tollis,
  %	{\em First-order phase transition in
  %   ferromagnetic/superconductor/ferromagnetic trilayers}
  Phys. Rev. B {\bf 69}, 104532 (2004).
	
\end{thebibliography}
\end{document}